\documentclass[aps,prb,reprint,a4paper]{revtex4-1}
\usepackage[utf8]{inputenc}
\usepackage[hidelinks]{hyperref}
\usepackage{amsmath}
\usepackage{graphicx}
\usepackage[english]{babel}
\usepackage[parse-numbers=false]{siunitx}
\usepackage{acronym}
\usepackage{nicefrac}
\usepackage{color}

\newcommand\sub[2]{{#1}^{\phantom\dagger}_{#2}}

\newcommand\cc[2]{{#1}^\dagger_{#2}}
\newcommand\ca[2]{\sub{#1}{#2}}
\newcommand\numop[2]{\cc{#1}{#2}\ca{#1}{#2}}

\newcommand\bra[1]{\left\langle#1\right|}
\newcommand\ket[1]{\left|#1\right\rangle}
\newcommand\HC{HC}

\newcounter{subfig}[figure]

\newacro{NRG}[NRG]{numerical renormalization group}
\newacro{SEM}{scanning electron microscopy}

\begin{document}

\title{Tuning Yu-Shiba-Rusinov States in a Quantum Dot}
\date{\today}
\author{Anders Jellinggaard}
\author{Kasper Grove-Rasmussen}
\author{Morten Hannibal Madsen}
\author{Jesper Nygård}
\affiliation{Center for Quantum Devices, University of Copenhagen}

\begin{abstract}
We present transport spectroscopy of sub-gap states in a bottom gated InAs
nanowire coupled to a normal lead and a superconducting aluminium lead. The
device shows clearly resolved sub-gap states which we can track as the coupling
parameters of the system are tuned and as the gap is closed by means of a
magnetic field. We systematically extract system parameters by using numerical
renormalization group theory fits as a level of the quantum dot is tuned through
a quantum phase transition electrostatically and magnetically. We also give an
intuitive description of sub-gap excitations.
\end{abstract}

\maketitle

\section{Introduction}

Hybrid superconductor-quantum dot devices\cite{de_franceschi_hybrid_2010} are
heavily employed in recent experimental programs. For instance, quantum dots
serve as an integral component of proposals to form\cite{sau_realizing_2012,
leijnse_parity_2012}, manipulate\cite{flensberg_non-abelian_2011,
leijnse_quantum_2011, AliceaNatPhys2011}, and probe\cite{leijnse_scheme_2011,
cao_probing_2012, liu_detecting_2011} Majorana bound states \cite{OregPRL2010,
SauPRL2010}. In Cooper pair splitters, the dynamics of quantum dots filter
local Andreev reflections from the desired non-local Andreev reflections to
form a source of entangled electrons.\cite{recher_andreev_2001,
hofstetter_cooper_2009, herrmann_carbon_2010}

In a dot-superconductor system, where the charging energy is larger than
the order parameter, quasiparticles in the superconductor bind to the dot by the
exchange interaction and give rise to sub-gap excitations.\cite{Kirsanskas2015arXiv}
When these quasiparticles form a singlet with electrons on the dot, the
resultant states are called Yu-Shiba-Rusinov states\cite{yu_bound_1965,
shiba_classical_1968, rusinov_theory_1969} and have historically been
investigated primarily through scanning tunneling
microscopy\cite{YazdaniScience1997, FrankeScience2011}. Only recently, have
these excitations been observed in transport
experiments.\cite{lee_spin-resolved_2014, Grove-RasmussenPRB2009,
pillet_andreev_2010, deacon_kondo-enhanced_2010, DirksNatPhys2011,
deacon_tunneling_2010, ChangPRL2013, KIMPRL2013, KumarPRB2014, SchindelePRB2014,
LeePRL2012, PilletPRB2013, DelagrangearXiv2016} We will give an intuitive
description of sub-gap excitations in the following section.

To experimentally investigate sub-gap excitations, we fabricated a bottom gated
normal metal/\allowbreak nanowire/\allowbreak superconductor device (N-NW-S),
which allows for the formation of a gate defined quantum dot proximitized to the
superconductor. The device shows clearly resolved sub-gap states that we can
track as the device is electrostatically tuned. In this way we follow a single
charge state of the dot, through the doublet to singlet quantum phase
transitions occurring as the barrier to the superconductor is lowered. We fit
measured excitations energies to a simulation developed for this purpose using
the non-perturbative \ac{NRG} method \cite{wilson_renormalization_1975,
bulla_numerical_2008, PilletPRB2013}, and in this way systematically extract
physical parameters of the device.

The system investigated is in many ways similar to N-NW-S devices where Majorana
bound states have been examined,\cite{MourikScience2012, das_zero-bias_2012,
AlbrechtNat2016} and a good understanding of the magnetic field behavior of
proximitized nanowire quantum dots is necessary to understand transport data of
these similar devices. We probe in detail the magnetic field behavior and
observe excitations apparently clinging to zero bias as the gap is about to
close, consistent with a recent experiment\cite{lee_spin-resolved_2014}.

\begin{figure}[htbp!]
  \includegraphics{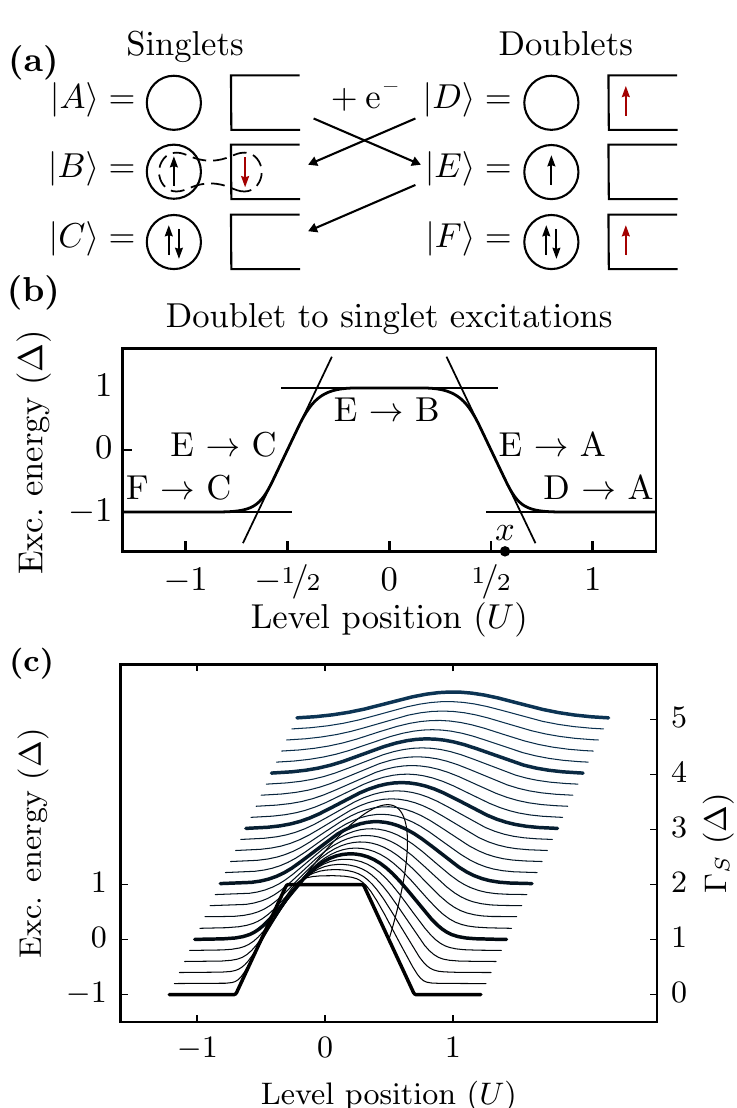}
  \caption{\label{fig:gamma_evolution}\label{fig:sub_gap_states}
    Excitations between states in a quantum dot/superconductor system.
    \textbf{(a)} The states under consideration with arrows in circles
    representing electrons in the dot and arrows in open rectangles
    representing bogoliubons. The dashed shape in the diagram for $\ket B$
    depicts a singlet correlation. The arrows annotated with +e$^-$ 
    show
    dominant sequential tunneling processes for transport going from N to S at
    the level position marked $x$ in (b).
    \textbf{(b)} A schematic diagram showing sub-gap excitations in the
    system, with and without anti-crossings induced by the coupling between
    the dot and superconductor.
    \textbf{(c)} \ac{NRG} simulations of the lowest doublet to singlet
    transitions for different values of the coupling density of states,
    $\Gamma_S$. For all curves in (c), we have $U=5\Delta$. The curve going
    across the traces mark an excitation energy of zero. Traces have been
    offset for clarity as indicated on the right-hand axis.
  }%
  \refstepcounter{subfig}%
  \label{subfig:states}%
  \refstepcounter{subfig}%
  \label{subfig:mixing}%
  \refstepcounter{subfig}%
  \label{subfig:gamma_evolution}%
\end{figure}

\subsection{Sub-gap states}\label{sec:subgapstates}

We consider a quantum dot described by the Anderson model (full hamiltonian in
Appendix.~\ref{sec:model}) with a single level at $\epsilon$ and a charging
energy of $U$, coupled to a normal lead and to a superconducing lead with
order parameter $\Delta$. The strength of the coupling to lead~$\alpha$
($\alpha = N, S$), is governed by the corresponding tunneling density of
states, $\Gamma_\alpha = 2\pi|t_\alpha|^2\nu_F$, where $t_\alpha$ is the
tunneling coefficient of lead~$\alpha$ and $\nu_F$ is the density of states of
lead~$\alpha$ near the Fermi level. Our data is collected in a regime where
$\Gamma_N$ is small compared to the other energies of the system, so we
consider the normal lead to be a tunnel probe\footnote{Recent numerical work
suggest that the normal lead may have a non-pertubative effect on the system,
so this approximation may not be entirely justified.\cite{zitko_shiba_2015}}
which is used to probe the quantum dot/superconductor system.

The nature of sub-gap excitations in such a system depends on the relative size
of $\Delta$ and $U$.\cite{Kirsanskas2015arXiv} If $\Delta$ is large, the system
can be understood in terms of repeated Andreev reflections giving rise to
Andreev bound states.\cite{MengPRB2009} When $U$ is large, Andreev reflections
are supressed, and instead we need to think in terms of quasiparticles
(bogoliubons) in the superconducting lead. We will here develop an intuition for
excitations in this case.

First, for vanishing $\Gamma_S$, we know exactly what the eigenstates of the
model are, and we will be focusing, in particular, on the states shown in
Fig.~\ref{fig:sub_gap_states}. These are the lowest energy singlets and doublets
(only half the doublets are shown) for different values of $\epsilon$. The
illustrated states are
\begin{align*}
  \ket A &= \ket0
  & \ket D &= \cc \gamma\uparrow \ket0 \\
  \ket B &=
    \tfrac1{\sqrt2} \bigl(
      \cc \gamma\downarrow \cc d\uparrow
      - \cc \gamma\uparrow \cc d\downarrow
    \bigr) \ket0
  & \ket E &= \cc d\uparrow \ket0 \\
  \ket C &= \cc d\uparrow \cc d\downarrow \ket0
  & \ket F &= \cc d\uparrow \cc d\downarrow \cc \gamma\uparrow \ket0,
\end{align*}
where we have defined $\cc \gamma\sigma$ as the operator that creates the lowest
energy bogoliubon\cite{tinkham_introduction_2004} with spin $\sigma$, and $\cc
d\sigma$ as the operator which creates an electron on the dot with spin
$\sigma$. The figure also shows the energy of relevant excitations between these
states.

As $\Gamma_S$ is turned up, the singlet states are mixed resulting in avoided
crossings, and the same happens for the different doublet states. For instance,
the coupling between $\ket A$ and $\ket B$ causes the excitation energy inside
the gap to move down towards the center of the gap. The other bogoliubons (those
of higher energy) will all move the sub-gap excitation in the same direction.

Eventually, this simple picture breaks down, as states with more than one
bogoliubon become a significant factor in forming the low energy eigenstates.
For higher $\Gamma_S$, it is not possible to find a simple theory that covers
the entire range of $\epsilon$ and lends itself to a clear physical
understanding, and one has to resort to numerical procedures. In this vein,
Fig.~\ref{fig:gamma_evolution}(c) shows the lowest energy doublet to singlet
excitation as a function of the level position and $\Gamma_S$ as found using
\ac{NRG} simulations. In the middle of the Coulomb valley the doublet to singlet
excitation energy decreases with increasing $\Gamma_S$, indicating a
stabilization of the singlet state, and eventually the energy crosses zero,
which is an example of a second order quantum phase
transition.\cite{lee_spin-resolved_2014, sachdev_quantum_2007} For larger
$\Gamma_S$, the ground state remains a singlet for all level positions, even as
the expected number of electrons on the dot changes by 2.

The \ac{NRG} method has been applied to the proximitized Kondo
model,\cite{satori_numerical_1992, sakai_numerical_1993} the proximitized
Anderson model,\cite{yoshioka_numerical_2000} and to the normal metal/quantum
dot/superconductor system\cite{tanaka_numerical_2007, zitko_shiba_2015} in the
literature, and generally recreates the features seen in real systems fairly
accurately, as our fits below also indicate. Note, that we are using a newly
developed \ac{NRG} program which does not exploit symmetries in
the system to speed up the algorithm.\footnote{A.\ Jellinggaard et al.\ (in
preparation).} Consequently, we only keep 160 states from each link of the
chain. In Appendix \ref{app:NRG} we compare results from our program to
the phase diagram in Ref.~\onlinecite{zitko_shiba_2015} to show that there is
reasonable agreement between the output from our program and that of an
established program running a simulation with more states retained.

\subsection{Transport}

We imagine that transport in the device is primarily sequential in electrons
tunneling from the N electrode to the dot--S system. This is possible once
states with different numbers of fermions are mixed. In
Fig.~\ref{subfig:states} we have tried to illustrate the dominant sequential
transport processes moving electrons from N to S when the level position is
near $\nicefrac12\,U$, i.e. at $x$ in Fig.~\ref{subfig:mixing}. In this case, considering again---artificially---only one
bogoliubon state, the lowest energy singlet, $\ket s$, is a linear combination
of primarily $\ket A$ but with some weight on $\ket B$ and $\ket C$, and the
lowest energy doublet, $\ket d$ consist mainly of $\ket E$ with some weight on
$\ket D$ and $\ket F$. Transport occurs by repeatedly swapping the state
between $\ket s$ and $\ket d$ by adding electrons to the dot from the N lead.

Fermi's golden rule tells us that the rate at which we go from $\ket s$ to
$\ket d$ is proportional to $\bigl|\bra d \cc d\uparrow \ket s \bigr|^2$
which, for low $\Gamma_S$, is close to what we would expect for a non-
proximitized dot. Going from $\ket d$ to $\ket s$ occurs at a rate
proportional to $\bigl|\bra s \cc d\uparrow \ket d \bigr|^2$, which is smaller
because only terms involving $\ket B$ and $\ket D$ or $\ket C$ contribute,
c.f. Fig.~\ref{subfig:states}. Intuitively, we have to move two electrons
across the barrier to S in this transport process.

\begin{figure*}
  \includegraphics{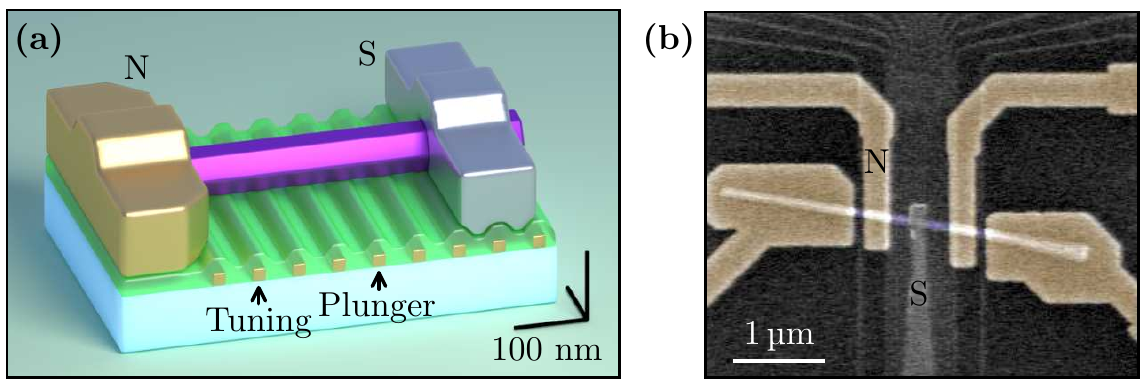}
  \caption{\label{fig:device}
    \textbf{(a)} Artist impression of a $\SI{0.6}{\um} \times \SI{0.4}{\um}$
    cutout of the device, to scale. The model shows the surface of the SiO$_2$
    substrate, bottom gates, insulating HfO$_2$ (shown in green), InAs
    nanowire, gold contact, and aluminium contact. Details are in
    Appendix~\ref{sec:fab}. We assign names to two of the gates as shown.
    \textbf{(b)} \acs{SEM} micrograph of a lithographically similar device.
    Note, that only the part of the device between the gold electrode, N, and
    the aluminium electrode, S, is used.
  }%
  \refstepcounter{subfig}%
  \label{subfig:dev_artist}%
  \refstepcounter{subfig}%
  \label{subfig:dev_sem}%
\end{figure*}

\begin{figure*}
  \includegraphics{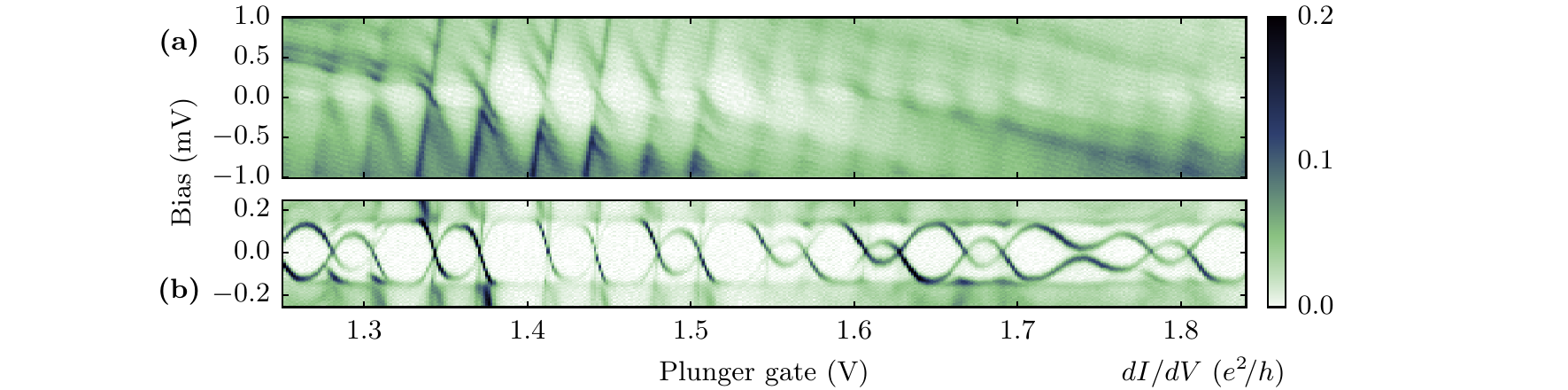}
  \caption{
    Differential conductance at 35~mK with and without an externally applied
    \SI{150}{mT} in-plane field. The field drives the aluminium contact normal
    in (b). Regions that show heavily tunnel-broadened Coulomb diamonds also
    show sub-gap excitations far inside the gap when the Al contact is
    superconducting.
  }%
  \refstepcounter{subfig}%
  \label{fig:intro_n}%
  \refstepcounter{subfig}%
  \label{fig:intro_s}%
\end{figure*}

\section{Experimental results}

The device is a bottom gated \SI{70}{nm} diameter InAs nanowire with one Ti/Au
contact and one Ti/Al contact approximately \SI{330}{nm} apart. The bottom gates
have a \SI{55}{nm} center-to-center distance and are separated from the nanowire
by a \SI{24}{nm} HfO$_2$ 
dielectric. The contacts are both well coupled to the
nanowire compared to the deliberate transport barriers we impose with the bottom
gates to form the dot, and the Ti/Al contact is superconducting with $\Delta =
\SI{0.14}{mV}$. Further fabrication details can be found in
Appendix~\ref{sec:fab}.

Figure~\ref{subfig:dev_artist} shows a scale model of our device and our
designation of a ``tuning gate'', $V_T$, and a ``plunger gate'', $V_P$. A
\ac{SEM} micrograph of a similar device is shown in Fig.~\ref{subfig:dev_sem},
where only the nanowire segment between the N and S electrodes is probed by
transport. In all plots, we apply a bias, $V_{sd}$ to the aluminium contact
and measure differential conductance, $dI/dV$, through the device at a
temperature of 35~mK. Figure~\ref{fig:intro_n} shows typical transport data
with the aluminium contact driven normal by a field, and
Fig.~\ref{fig:intro_s} shows corresponding bias spectroscopy at zero field
where the superconducting gap is visible as a horizontal band of low
differential conductance between $V_{sd} = -\SI{0.14}{mV}$ and
$\SI{0.14}{mV}$. The normal state data shows the usual Coulomb diamonds for
$V_P < \SI{1.6}{V}$, but for $V_P > \SI{1.6}{V}$ these diamonds become
difficult to resolve, as the excitations are heavily tunnel broadened by the
coupling the aluminium contact. In regions where the excitations are broadened
in the normal state data (cf.\ Fig.\
\ref{fig:intro_n}), which we attribute to a strong coupling to the aluminium
contact, we see that the sub-gap excitations in Fig.~\ref{fig:intro_s} are
pushed far inside the gap. In the remainder of this article, we investigate
how these sub-gap excitations respond to gate tuning and to small (less than
$B_c$) magnetic fields.

\subsection{Gate tuning}

\begin{figure}
  \includegraphics{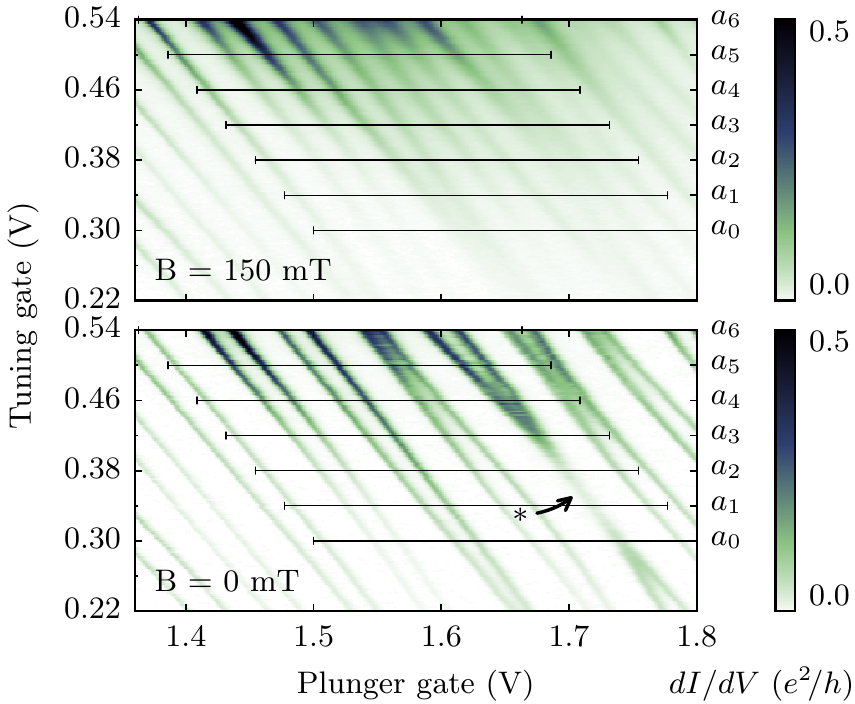}
  \caption{\label{fig:stability}
    Conductance at zero bias as a function of the tuning gate and the raw
    plunger gate potential, with and without a magnetic field driving the
    aluminium contact normal. The lines in these plots shows the cuts done by
    $a_0$-$a_6$ of Fig.~\ref{fig:tune}. For certain configurations of the tuning
    gate, the ground state remains a singlet as a dot level is brought past the
    fermi-level with the plunger gate, and this is evident at the
    \textasteriskcentered.
  }
\end{figure}

\begin{figure}
   \includegraphics{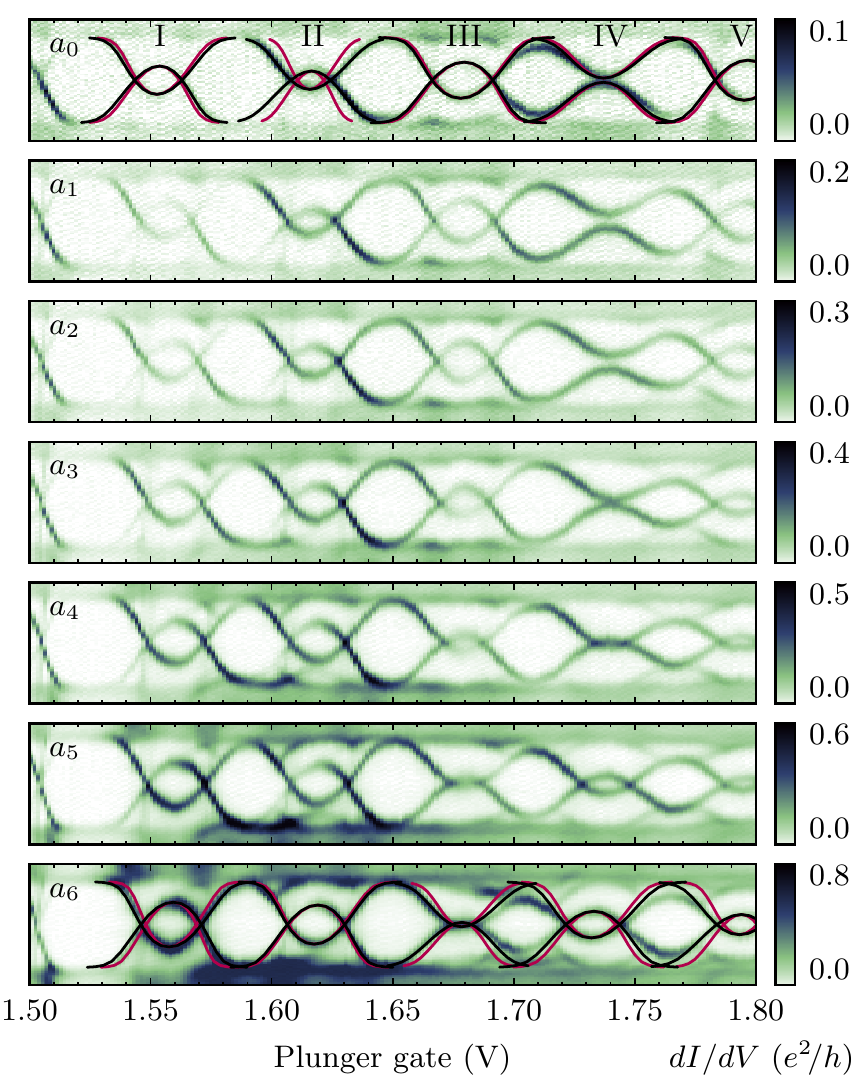}
  \caption{\label{fig:tune}
    Bias spectroscopy of the sub-gap states for different gate configurations.
    In every plot, the y-axis is the potential of the superconducting lead
    relative to the normal lead and ranges from \SIrange{-0.2}{0.2}{mV}. Note
    the different $dI/dV$ scales. In the plot $a_n$, the tuning gate is set to
    \SI{(300 + 40n)}{mV}. We adjust for cross capacitance as described in the
    text. Example \ac{NRG} fits are overlaid plots $a_0$ and $a_6$. The red
    curves are made using Method~1 and the black curves are made using Method~2,
    see text for details.
  }
\end{figure}

Figure~\ref{fig:stability} shows the zero bias differential conductance of the
device as a function of the potential, $V_P$ and $V_T$, of the plunger and
tuning gate. Both gates couple to the dot and have capacitances of $C_P \approx \SI{5.5}{aF}$ and $C_T \approx \SI{3}{aF}$ (lever arms $\alpha_P \approx 0.035$ and $ \alpha_T \approx 0.02$) for the plunger and tuning gates, respectively.
We define $V_P'= V_P + (V_T - \SI{0.3}{V}) \times 0.57$ to compensate for this
cross capacitance, and will use this for all subsequent figures instead of
$V_P$. Note the overall increase in conductance for increasing $V_T$, which we
ascribe to a lowering of the barrier to the normal lead consistent with the
position of the tuning gate. Later, we shall see that $\Gamma_N$ depends on
$V_T$ exponentially, which supports this assertion. Also evident in these
plots is a quantum phase transition (at the
\textasteriskcentered), which will become clearer in later plots.

Figure~\ref{fig:tune} shows how the sub-gap excitations respond to tuning, and a
few trends are apparent going from low ($a_0$) to high ($a_6$) $V_T$. First, we
see again the overall increase in conductance with higher $V_T$. Secondly, as
$V_T$ is changed, the sub-gap excitations of Fig.~\ref{fig:tune} shift in
energy, with no overall trend, which we interpret as mesoscopic fluctuations of
$\Gamma_S$ as the wavefunctions of the dot states are perturbed by the changing
$V_T$ and $V_P$. The quantum phase transition is very clear in this plot,
occurring for the charge state labeled IV around $V_T = \SI{0.44}{V}$, i.e.\
between $a_3$ and $a_4$. The other charge states do not undergo this kind of
quantum phase transition in the data shown. We point out, that the transitions
are not significantly tunnel-broadened compared to the size of the gap, so we
can assume that normal lead is weakly coupled to the dot.

\begin{figure}
    \includegraphics{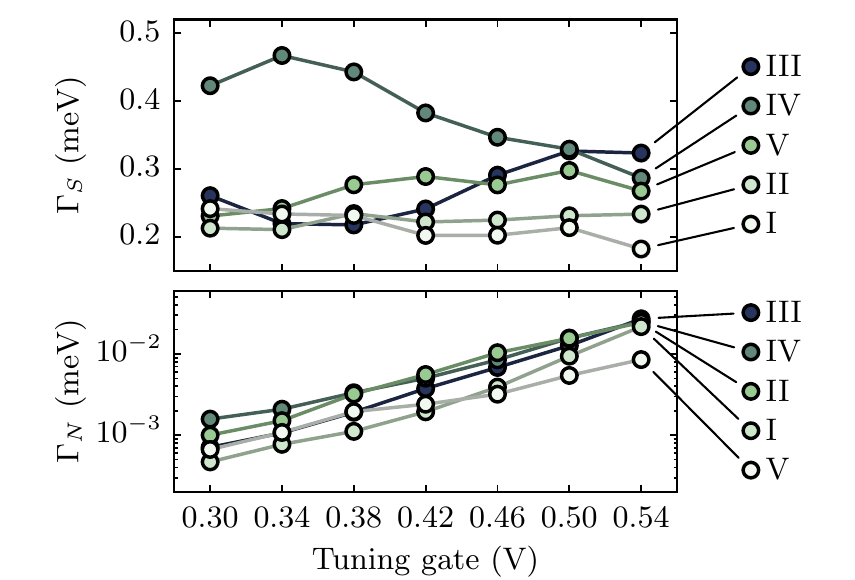}
  \caption{\label{fig:gammas}
    Coupling strengths $\Gamma_S$ and $\Gamma_N$ of each level transition shown
    in Fig.~\ref{fig:tune}. We extract $\Gamma_S$ from our \ac{NRG} fits, and
    use the conductance on resonance in the $B=\SI{150}{mT}$ data of
    Fig.~\ref{fig:tune} to find $\Gamma_N$. There is one trace in the plot for
    each level transition, and the roman numerals refer back to the labels in
    Fig.~\ref{fig:tune}.
  }
\end{figure}

To extract quantitative parameters for the system, we fit a model to the data
consisting of single levels independently interacting with the superconductor,
such that each level is described by the proximitized Anderson model. In this
model, each level is described by the following parameters: The charging
energy $U$, the potential of the plunger gate at the center of the
corresponding Coulomb valley $V_0$, the plunger gate capacitance $C_P$, and
the coupling strengths $\Gamma_S$ and $\Gamma_N$. We will find quantitative
estimates for all these parameters.

Specifically, we find $U$ from the height of the corresponding Coulomb diamond
in Fig.~\ref{fig:intro_n}, and we find $V_0$ by looking at
Fig.~\ref{fig:tune}. We initially assume $\Gamma_N$ is weak, in which case it
has little effect on level positions and does not drive the NW-S system out of
equilibrium, and we find $C_P$ and $\Gamma_S$ using one of two methods both
involving a fit based on the \ac{NRG} method: for Method-1, we find $C_P$ from
the normal state data in Fig.~\ref{fig:stability} and use $\Gamma_S$ as a
fitting parameter to fit the observed level positions. For Method-2, we use
both $C_P$ and $\Gamma_S$ as fitting parameters. Fits to two of the datasets
are shown in Fig.~\ref{fig:tune} for both methods, the rest are included in
the supplementary information.

Generally both methods reproduce the gate dependence of the sub-gap state
excitations well. The most significant divergence is around $V_P' =
\SI{1.70}{V}$ and $V_P' = \SI{1.76}{V}$, where additional excitation lines are
present inside the gap. The presence of these lines suggest that the levels are
not independent in this region.

Having found the values of $\Gamma_S$ at each level crossing from our \ac{NRG}
fits, we extract $\Gamma_N$ from the conductance at each Coulomb peak when the
superconductor is driven normal by an external magnetic field, i.e. from the
data in Fig.~\ref{fig:stability}. Specifically
\begin{equation}\label{eq:gpeak}
  G_{peak} = \frac{e^2}h \frac
    {4\Gamma_S\Gamma_N}
    {(\Gamma_S + \Gamma_N)^2}
  ,
\end{equation}
where $G_{peak}$ is the maximal conductance of the device at the Coulomb
peak.\cite{jorgensen_critical_2007} The values of $\Gamma_N$ and $\Gamma_S$
that we extract are shown in Fig.~\ref{fig:gammas}. $\Gamma_N$ shows an
exponential dependence on the tuning gate potential, as expected for an
electron tunneling through a potential barrier. In contrast, $\Gamma_S$ varied
non-monotonically and did not have a systematic dependence on gate. Therefore,
we attribute the variations we do see in $\Gamma_S$ to mesoscopic fluctuations
caused by perturbations of the dot wavefunctions, rather than a changing
potential barrier.

\subsection{Results: Behavior at Field}

\begin{figure}
  \includegraphics{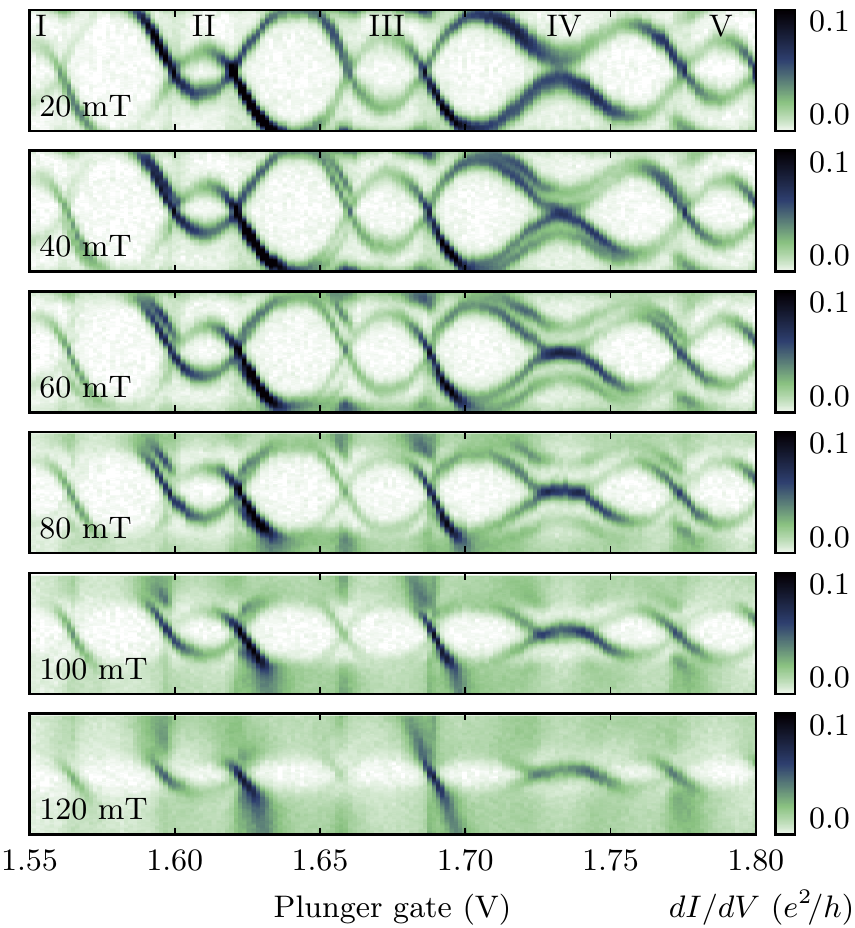}
  \caption{\label{fig:field}
    Evolution of the sub-gap states as the gap closes upon the application of an
    external field in plane with the substrate perpendicular to the nanowire.
    The tuning gate is set to $V_T = \SI{0.34}{V}$ in all these plots, and the
    bias range is $\pm\SI{125}{\micro V}$. The roman numerals on odd charge
    states refer back to the labels in Fig.~\ref{fig:tune}.
  }
\end{figure}

\newcommand\bvfcap{\label{fig:bias_vs_field}
  Magnetic field dependence of sub-gap transport in the center of a level
  transition for different values of the tuning gate potential. In the data sets
  $b_0$ through $b_5$ the field was applied in the plane of the sample
  perpendicular to the nanowire, and in the data sets $c_0$ through $c_5$ the
  field was held at \SI{60}{mT} and rotated in the plane of the sample with a
  direction of 0 radians being perpendicular to the wire. The data has been
  corrected for a drifting zero-bias across the device as detailed in the
  supplementary information. The plots also show data from an \ac{NRG}
  simulation, specifically the allowed excitations from the ground states of the
  system (red lines). The black lines show a phenomenological model of the gap
  closing used as an input to the simulation. Additional input to the simulation
  includes the $\Gamma_S$ values from Fig.~\ref{fig:gammas} and a g-factor found
  by fitting the plot $c_1$ at 0 radians.
}

  \begin{figure*}
      \includegraphics{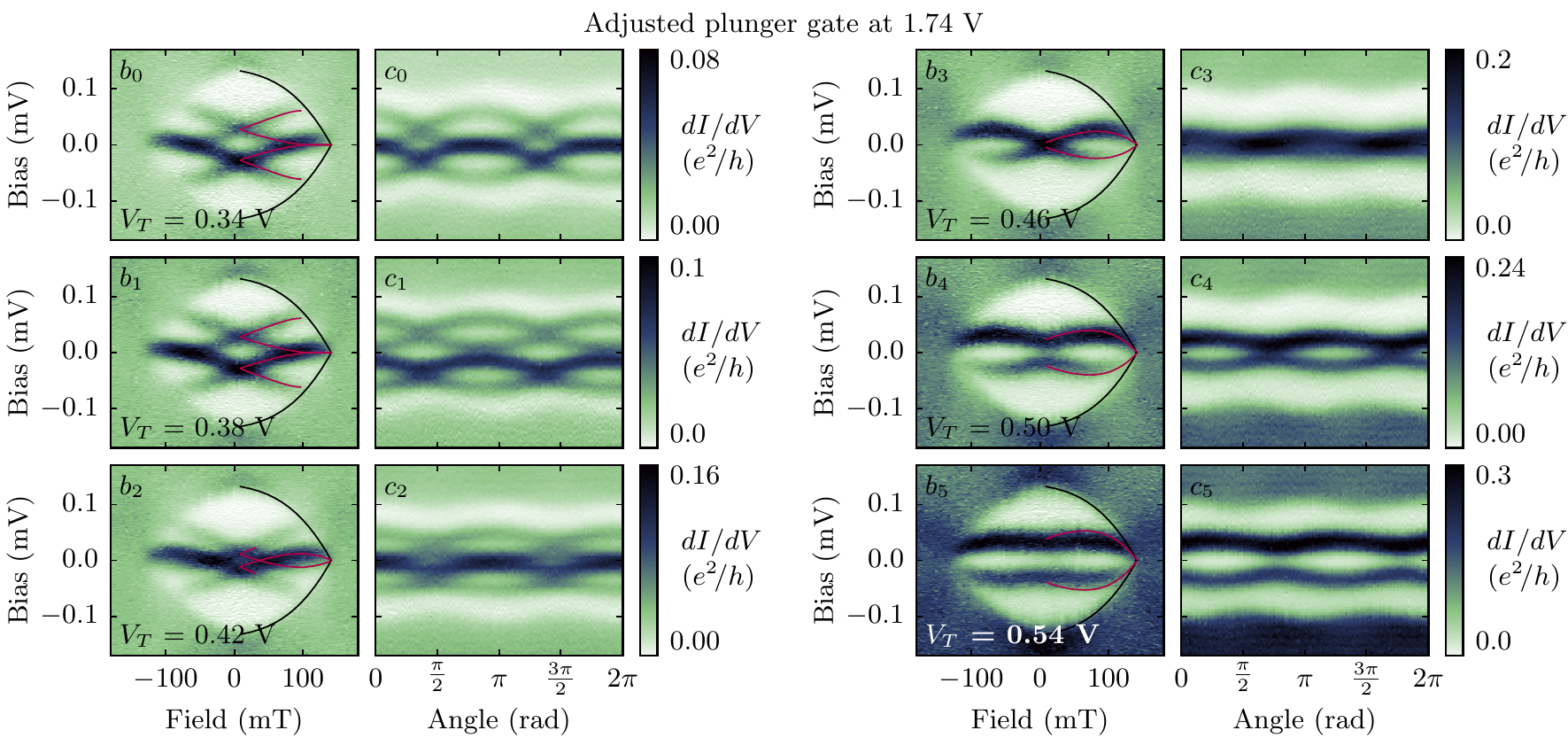}
    \caption{\bvfcap
    }
  \end{figure*}

We now turn to the magnetic field dependence of the sub-gap states.
Figure~\ref{fig:field} shows what happens to the plot $a_1$ in
Fig.~\ref{fig:tune} as a field is applied in the plane of the sample in a
direction perpendicular to the nanowire. As the field increases, the doublet
states Zeeman split, which is clear where the ground state is a singlet. When
the ground state is a doublet, only one excitation is possible from the ground
state, and only one peak is seen in transport.\cite{lee_spin-resolved_2014} When
analysing our data, we will augment the Anderson model hamiltonian of the dot
from the introduction with a Zeeman term of the form
\begin{equation}
  H_Z = g \mu_B B \cdot S,
\end{equation}
where $g$ is the g-factor of the level, $\mu_B$ is the Bohr magneton, $B$ is the
strength of the magnetic field, and $S$ is the spin of the dot; note that we
always align the $z$-axis of the spin basis with the magnetic field. In our
quantum dot, the effective g-factor varies significantly between
levels,\cite{CsonkaNaneLett2008} and even within a single level. For instance,
in the charge state labeled II in Fig.~\ref{fig:field}, the splitting of the
excitations line left of center (near more negative $V_P$) is very different
from the splitting right of center.

At the charge state labeled IV in Fig.~\ref{fig:tune} we are able to induce a
quantum phase transition by a applying a magnetic field, so we focus on this
level crossing. Figure~\ref{fig:bias_vs_field} shows the dependence of transport
at the center of the crossing both as the field magnitude is increased and as
the field is rotated. As in our other dataset, we again note the absence of a
transition from the excited member of the doublet to the singlet, which is what
causes the peaks in Fig.~\ref{fig:bias_vs_field} for $V_T > \SI{0.42}{V}$ to
only move in one direction instead of splitting.

As is apparent in the plots $c_0$ through $c_5$ of Fig.~\ref{fig:bias_vs_field},
the g-factor of our system shows a high degree of anisotropy. This is a common
property of quantum dots in InAs nanowires \cite{CsonkaNaneLett2008,
schroer_field_2011} and was also addressed by Lee et al.\ for YSR
states.\cite{lee_spin-resolved_2014}

In the bias vs.\@ field strength plots of Fig.~\ref{fig:bias_vs_field},
specifically plots $b_0$, $b_1$, and $b_2$, we note that the excitation of the
doublet that moves down in energy has an apparent tendency to stick to zero
bias. This effect has been observed before and can be understood in terms of a
level repulsion from the gap states as the gap closes, pinning the excitations
near zero energy.\cite{lee_spin-resolved_2014}

We estimate the level positions from the data plotted in $c_1$ of
Fig.~\ref{fig:bias_vs_field}, and fit the g-factor at angles of \SI{0}{\radian},
\SI{0.9 \pi}{\radian}, and \SI{1.4 \pi}{\radian} using our \ac{NRG} model. The
latter two angles correspond to minimal and maximal Zeeman splitting, note
that the splitting at \SI{0.9 \pi}{\radian} is hard to estimate precisely. For
these angles, we find g-factors of approximately 22, 8, and 23 respectively.

We use the g-factor at \SI{0}{\radian} along with the values of $\Gamma_S$
found for each tuning gate value earlier, to simulate how the states split
with applied field, i.e. to recreate the level positions seen in the plots
$b_0$ through $b_5$. The resulting level transitions are plotted in the figure
and show good agreement with the data. We plot excitations from the ground
state only, but in the plot $b_2$, transport is also possible from the doublet
state, presumably because the doublet is thermally excited.

\section{Conclusion}

The device presented in this paper had two features that complement each other:
Transparent contacts and well coupled bottom gates with a large admissible
voltage range. This made it feasible to make completely gate defined contact
barriers in the device, and tune coupling parameters over a large range while
keeping mostly single-dot behavior. In combination with well resolved sub-gap
states, the device provided an excellent platform to study the dependence of
Yu-Shiba-Rusinov states to $\Gamma_S$-tuning and to magnetic fields. Future
studies may involve testing the recent theoretical predictions that the
singlet-doublet phase diagram is modified by the normal metal
coupling.\cite{zitko_shiba_2015}

For the data presented in this paper, we used a gate between the quantum dot and
the normal contact to tune our device. This had a large effect on $\Gamma_N$
which in turn has only a small effect on level positions. On the other hand,
mesoscopic fluctuations of $\Gamma_S$ (on the other side of the device) caused
by this tuning, has a large and, a-priori, unpredictable effect on $\Gamma_S$.
Effects like this can appear in gated quantum dot devices, whether it
involves a superconducting contact or not, but this device is an interesting
example as the two contact barriers influence transport in very dissimilar ways.

Modeling the device using the proximitized Anderson model by means of the
\ac{NRG} method yielded excitation energies in good agreement with our data, and
the coupling parameters extracted from these fits follow the potentials of the
bottom gates in a physically reasonable way. The behavior under magnetic field
is entirely consistent with a simple Zeeman splitting in combination with the
gap closing. This behavior has been described
before.\cite{lee_spin-resolved_2014} However, here we model this scenario
quantitatively using the \ac{NRG} method starting from parameters determined at
zero field, and show good agreement with observed data.

\begin{acknowledgments}

  We would like to thank Ramon Aguado and Jens Paaske for helpful discussions
  and Peter Krogstrup, Claus B. Sørensen and Erik Johnson for experimental
  contributions. Funding for this project was provided by the EU FP7 project
  SE2ND (Source of Entangled Electrons in Nano Devices), the Carlsberg
  Foundation, the Danish Research Council DFF-FNU and the Danish National
  Research Foundation.

\end{acknowledgments}

\appendix

\section{Fabrication details}\label{sec:fab}

The bottom gates were fabricated on a Si substrate with \SI{500}{nm} oxide, and
are composed of \SI{5}{nm} Ti and \SI{12}{nm} Au. These gates have a center to
center distance of \SI{55}{nm}. The gates are covered with \SI{24}{nm} HfO$_2$
deposited by atomic layer depositioning. This HfO$_2$ is deposited in three
\SI{8}{nm} layers of successively smaller extent to avoid \emph{fencing}, where
the oxide does not break off cleanly where it meets resist walls and instead
stand proud off the surface after lift-off.

\SI{70}{nm} diameter InAs nanowires were deposited from a suspension in
isopropanol. In the evaporation chamber, immediately prior to metalization of
each contact, argon ion milling was used to remove the native oxide from the
nanowire. The Au contact uses a \SI{10}{nm} Ti sticking layer, and the Al
contact uses a \SI{5}{nm} Ti sticking layer. Compared to the data shown in
this article, the device is significantly more conductive when a higher
potential is applied to the bottom gates, suggesting that the tunnel barriers
seen in the data are gate defined as opposed to contact defined.

\begin{figure}
  \includegraphics{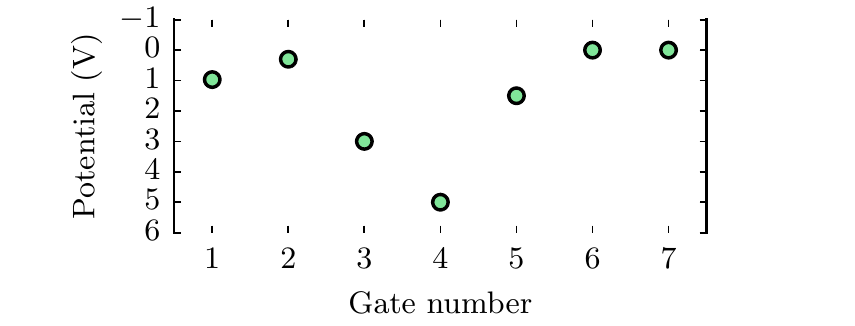}
  \caption{\label{fig:tuning}
    Electric potentials applied to each relevant bottom gate. The gates are
    numbered starting from the gold side of the device (left side in
    Fig.~\ref{fig:device}). Gate nr.~2 we call the tuning gate and gate nr.~5,
    which is more strongly coupled to the energy levels of the dot, we call
    the plunger gate.
  }
\end{figure}

Tuning the potentials of the gates allow the device to be operated in
different regimes; in this paper we focus on single dot behavior by forming a
central potential dip (see Fig.~\ref{fig:tuning}). We note that bottom gates
under a contact, for instance the second bottom gate from the right in
Fig.~\ref{fig:device}, generally do not show any significant effect on
transport through the device. This suggest that the gates are strongly
screened, or that the contacts---by diverting current out of the wire already
very near the edge of the contact---make it a moot point whether the sections
of nanowire above these gates are depleted or not.

The device investigated in this paper forms part of a larger two-sided device.
To avoid complications from the other side of the device, this part of the
nanowire was electrostatically depleted during measurements.

\section{Model details}\label{sec:model}

For the discussions in the article and in our models, we use the following
hamiltonian
\begin{equation}
  H = H_d + H_S + H_{t\textsc s} + H_{t\textsc n},
\end{equation}
with each part given below. The quantum dot has the hamiltonian
\begin{equation}
  H_d = \sum_\sigma \sub\epsilon\sigma \numop c \sigma
      + \frac U2 \Big( \sum_\sigma \numop c \sigma - 1 \Big)^2
\end{equation}
where $\cc c \sigma$ creates an electron with spin $\sigma$ on the dot, $U$ is
the charging energy of the dot, and $\epsilon$ is the the single particle
energy of the dot. The dot is coupled to the two leads by
\begin{align}
  H_{t\textsc s} &= \sum_{k\sigma} \sub t{\textsc s}
      \cc c \sigma \ca c {k\sigma} + \HC
  \\
  H_{t\textsc n} &= \sum_{k\sigma} \sub t{\textsc n}
      \cc c \sigma \ca f {k\sigma} + \HC,
\end{align}
where $\cc c {k\sigma}$ creates an ordinary fermion in the superconducting
lead with momentum $k$ and spin $\sigma$, $\cc f {k\sigma}$ creates a
quasiparticle in the normal lead, and the $t$'s are tunneling coefficients
assumed spin and momentum independent. The hamiltonian of the superconductor
is
\begin{equation}
  H_S = \sum_{k\sigma} \xi_{k\sigma}^{\phantom\dagger}
      \ca c{k\sigma}\cc c{k\sigma}
    + \sum_k
      \left(
        \Delta \ca c{k\uparrow} \ca c{-k\downarrow}
      + \Delta \cc c{-k\downarrow} \cc c{k\uparrow}
      \right)
\end{equation}
where $\Delta$ is the order parameter of superconductor which we assume is
real.

For our \ac{NRG} simulations, we assume that $\sub t{\textsc n}$ and the
temperature of the system are negligible, we discretize the leads
logarithmically using a discretization factor, $\Delta$, of 2.5, and map the
system to a chain of fermions starting with the quantum dot. We add sites of
the chain one at a time, and at each step retain the lower 160 eigenstates.
The sub-gap excitation energies converge quickly,\cite{satori_numerical_1992}
so we only extend the chain to 25 sites. To simulate the gap closing with
applied field, we created, by hand, a table of gap size as function of applied
field based on the data in Fig.~\ref{fig:bias_vs_field}.

When calculating the peak conductance with the superconductor driven normal,
in Eq.~\eqref{eq:gpeak}, we assume $k_BT \ll
\Gamma_S + \Gamma_N$ consistent with our findings, and use a result from the
supplementary information of Ref.~\onlinecite{jorgensen_critical_2007}.

\section{Evaluating the NRG program}
\label{app:NRG}
\begin{figure}
  \centering
  \includegraphics{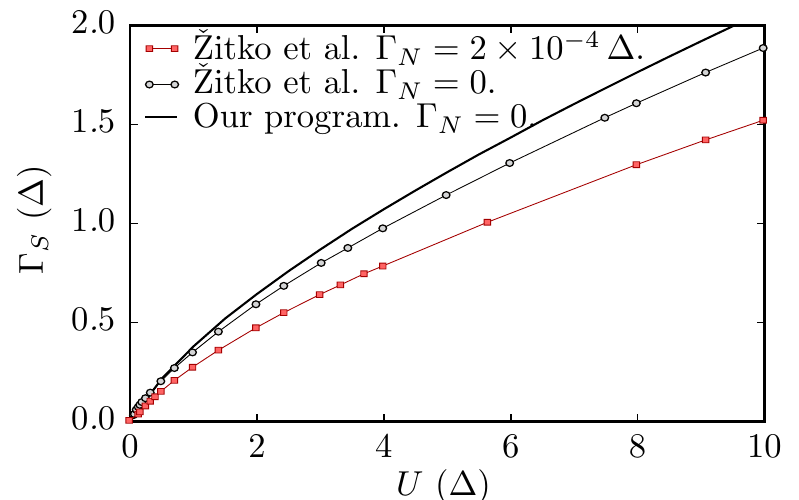}
  \caption{
    \label{fig:zitko}
    This figure compares output from our NRG program to Fig.~1 in
    Ref.~\onlinecite{zitko_shiba_2015}.
  }
\end{figure}

Since we are not exploiting symmetries in our NRG program, we cannot keep as
many states as others do in the calculations. Therefore we compared the output
of our program to a plot in \v Zitko et al, Physical Review B \textbf{91},
045441 (2015) (our Ref.~\onlinecite{zitko_shiba_2015}) showing the quantum
phase transition in a proximitized dot (see Fig.~\ref{fig:zitko}).
\v Zitko et al.\ also include simulations for a small coupling to the normal
lead, which we reproduce in Fig.~\ref{fig:zitko} to show that the error made by using our
program is small compared to the error made by not considering finite
$\Gamma_N$. Note, that our definition of $\Gamma_S$ differs from the one used
in \v Zitko et al.\ by a factor of two, so we scaled ours for this plot.

%

\end{document}